\documentclass[twocolumn, amsmath]{revtex4}

\usepackage{dcolumn}
\usepackage{bm}
\usepackage{graphicx}
\usepackage{color}
\usepackage{subfigure}
\usepackage{hyperref}
\usepackage{latexsym}
\usepackage{amsthm}
\usepackage{amssymb}
\usepackage{multirow}
\usepackage[english]{babel}
\usepackage{comment}

\usepackage{xcolor}
\usepackage{mdframed}
\mdfdefinestyle{mystyle}{
    backgroundcolor=gray!5
}

\usepackage{etoolbox}
\makeatletter
\patchcmd{\@verbatim}
  {\verbatim@font}
  {\verbatim@font\small}
  {}{}
\makeatother

\newcommand{\filippo}[1]{#1}
\newcommand{\jb}[1]{#1}

\begin{document}


\title{Quantifying \filippo{perceived} impact of scientific publications}

\author{ Filippo Radicchi}
\affiliation{Center for Complex Networks and Systems Research, School
  of Informatics and Computing, Indiana University, Bloomington, IN
  47408, USA}
  \email{filiradi@indiana.edu}

\author{Alexander Weissman}
\affiliation{Center for Complex Networks and Systems Research, School
  of Informatics and Computing, Indiana University, Bloomington, IN
  47408, USA}

\author{Johan Bollen}
\affiliation{Center for Complex Networks and Systems Research, School
  of Informatics and Computing, Indiana University, Bloomington, IN
  47408, USA}


\begin{abstract}
    Citations are commonly held to
    represent
    scientific impact.
    To date, however, there is no empirical 
    evidence in support of this postulate that is central to research assessment
    exercises and Science of Science studies.
    Here, we report on the first empirical verification of the degree to which 
    citation numbers represent 
    scientific impact 
    \filippo{as \jb{it is} actually perceived by experts in their respective field.}
    We run a large-scale survey of about
    $2,000$ corresponding authors who performed a pairwise impact assessment 
    task across more than $20,000$ scientific articles.
    Results of the survey show that 
    citation data and 
    \filippo{perceived}
    impact do not align well, unless
    one properly accounts for strong psychological biases 
    that affect \jb{the opinions of experts} with respect to their own papers vs.~those of others.
    First, researchers tend to largely
    prefer their own publications to the most cited papers in their field of research. Second, 
    there is only a mild positive 
    correlation between the number of citations of top-cited papers
    in given research areas and  
    expert preference in pairwise comparisons. 
    This also applies to pairs of papers with several orders of magnitude differences in
    \jb{their} total number of accumulated citations. However,
    when researchers were asked to choose among pairs of their own papers, 
    thus eliminating the bias favouring one's own papers over those of others,
    they did systematically prefer the most cited article. We conclude that, when scientists have full information and are making unbiased choices, expert opinion on impact is congruent with citation numbers.
\end{abstract}

\maketitle

\section{Introduction}
Metrics based on bibliographic data increasingly inform important decision-making
processes in science, such as  hiring, tenure, promotion, and 
funding~\cite{harzing2010publish, bor2006, bornmann2008does, 
lovegrove2008assessment, hornbostel2009funding, bornmann2011multilevel}. 
Many, if not most, of these bibliometric indicators are derived from article-level citation counts~\cite{bar2008informetrics,  van2010metrics},
e.g.~the h-index~\cite{hirsch2005index} and the journal 
impact factor~\cite{garfield2006history}, due to the wide availability of extensive bibliographic records. However, they rest on the frequently undeclared assumption that citations are an accurate and reliable reflection of scientific impact. Bornmann {\it et al.}~\cite{bornmann2008citation} and Sarigol {\it et al.}~\cite{sarigol2014predicting} find that citations can serve as reasonable approximations of social indicators such as popularity or success, but \jb{they do not support the most important assumption underlying most work in the area of citation-based impact 
indicators}, namely that \emph{citations quantify scientific impact}. 
Here, we investigate this assumption directly by determining whether citation data truly reflect scientific impact as it is perceived 
by expert scholars in their respective fields.
Such an assumption 
is central in the long-standing 
debate about the proper use of
citation data~\cite{Garfield1979}.
One common argument against the use of citation data in assessment exercises rests in particular on the doubts about their validity as indicators of true scientific impact~\cite{macroberts1996problems, adler2009, macroberts2010problems}. 
In fact, 
the literature frequently confounds ``impact'', ``influence'', and ``rank''
with citation-derived metrics~\cite{wuchty2007increasing, radicchi2008universality, wang2013quantifying} although 
these metrics can vary along multiple 
distinct dimensions~\cite{citations:bornman2008,bollen2009principal}.

Determining whether citations actually indicate scientific impact is an empirical question which can not be resolved by theoretical discussions of the perceived or presumed benefits {\it vs}. demerits of citation data alone. 
\filippo{
For this reason, we decided to define and empirically
quantify a novel post-publication metric, namely impact as perceived by the authors themselves. Our goal is to understand whether citation numbers ``truly'' reflect  a particular dimension of impact:  the influence or importance of papers in the daily practice of researchers.
}

We designed a large-scale survey at
\url{bigscience.soic.indiana.edu} to collect 
responses from thousands of 
experienced scholars in a multitude of 
disciplines (Fig.~\ref{figSURVEY}). 
We asked researchers to make a pairwise decision with regards to their
preference of one paper over the other. These decisions were made under 
two conditions: whether the articles were written by the expert scholars themselves or by other scholars. 
We then aggregated results over the entire population of respondents to
quantify the degree of correlation between the
pairwise preferences of respondents (i.e., perceived impact) 
and the actual difference in the number of citations accumulated 
(i.e., citation impact) for the pair of papers.
Our results indicate that, from 
the perspective of individual researchers, their assessment of 
impact \jb{vs.}~impact judged from citation data are \emph{only} 
related for pairs of their 
own papers. Every time that
a paper not co-authored by the individual is involved 
in the estimation 
of perceived impact, this comparison shows null or 
negative correlation with citation impact.

\section{Methods}
To build the infrastructure needed for the survey,
we took all scientific articles with a publication year up to 
$2013$ from the Web of Science (WoS) database. 
We associated every article with the total number of citations 
accumulated until $2013$ in the 
WoS citation network as an indication of its citation impact. 
We identified all articles that were associated with 
corresponding author(s) from three major public universities 
in the US: (1) Indiana University (IU), 
(2) University of Michigan (UMICH), 
and (3) University of Minnesota (UMN).
We chose IU to run the first pilot experiment since it is our home university \jb{; this would make initial data validation more straightforward}. UMICH and UMN were selected 
since they have the largest number of recent publications among all 
public universities in the US. All three universities 
host departments in almost all disciplines
of the natural and social sciences, offering a relatively unbiased sample 
of participants in terms of field of expertise. 
Articles were matched to their corresponding author(s) 
using the email address(es) provided
in the article metadata. For example, 
email addresses ending in ``indiana.edu'' 
were taken as an indication that the authors were based at IU.
Similarly for UMICH and UMN, we identified researchers
from those institutions by retaining email addresses
that ended with ``umich.edu'' and ``umn.edu,'' respectively.

In our data set, articles published prior
to 1995 did not provide author email addresses. 
After 1995, the proportion of articles carrying at least one email 
address increased rapidly each year. The vast majority of recent articles provide at least one email address. 
This may introduce a potential source of bias in our study towards relatively 
recent publications, but this procedure generated a number of 
very important advantages. First, the association
between articles and physical persons was virtually
free from \jb{homonymy-induced errors}. 
Second, and very 
important for our purpose, email addresses 
allowed us to  directly contact researchers and invite them to participate in our large-scale survey.

We sent 
an email message that contained a unique and customized URL 
to every potential participant (Fig.~\ref{figLetter}).
This URL pointed to a web page in which the respondent 
was presented with a maximum of ten pairwise comparisons between
scientific papers. In each comparison, we provided
only the journal and year of publication of the papers, \filippo{their title} and 
list of authors. For every pair of papers, 
the respondent was asked to select the one article she/he believed 
to be more ``influential'' for her/his own research.
Note that this task did not involve any consideration of, nor information 
on, citation data for any of the two papers. We tailored the survey for every respondent; papers were selected from three different pools that were constructed using information from the publication and citation record of the respondent: 
(i) ``Own'' publications (OWN), (ii)  ``Top cited'' articles (TCD), and (iii)  ``Random'' papers (RND).
The OWN pool consisted of articles written by the respondents themselves, i.e., they were associated with the email address of the responding author. The pool of TCD articles was constructed as follows.  First, we identified all articles appearing at least once in 
the reference lists of publications by 
the respondent. We eliminated from this set articles 
that appeared also in the pool of OWN articles. 
We then ranked the remaining articles based on their
citation impact, and selected the top 10
articles in the list. This procedure allowed us 
to populate the TCD pool with the most popular articles in the respondent's
area of research, but that were not written by the respondent her/himself.
The  pool of RND papers was simply generated as the union of 
articles (co-)authored or cited (not just the top cited) by 
all potential participants, thus comprising mainly articles unknown to the 
respondent. This pool was used only for the IU sample as a control set 
to check for the presence of possible systematic biases in the survey. 
Several respondents were confused when presented with
random papers. We therefore decided to remove the RND pool
from subsequent surveys made at UMICH and UMN. 
Although the IU respondents reported discomfort about having to make selections from the RND pool of papers, the data generated was useful to validate
construction of the other two pools, and to test for the absence of systematic 
biases in the visual format used in the online survey (Fig.~\ref{figSM1}). 
Once the three article pools were generated,  every comparison 
presented to respondents in their personalized survey was composed
of pairs of papers taken at random from the
various pools. This allowed us to collect
information about the preferences of respondents
among pairs of papers within the same pool and across different pools. We recorded the preferences expressed by every participant, 
and we used it to estimate the perceived impact of one paper with respect to 
the other in the comparison.

We sent emails for participation to a total of $19,546$ researchers (Table~\ref{tabSM1}). $1,819$ scholars participated in the survey, 
for a total of $12,000$ pairwise comparisons among $20,661$
distinct articles.
This is a low but acceptable response rate for an online survey with a self-selected sample~\cite{JCC4:JCC403}. Importantly, we didn't observe any systematic bias in the pool of participants in terms of academic age (Fig.~\ref{figSM2}), although we can not exclude the presence of all selection bias in  the sample of researchers who participated in the survey~\cite{bethlehem2010selection}.
We remark that our estimate of the response rate is conservative; some of the email addresses we used to contact researchers might be no longer active, e.g., through retirement or change of affiliation.
\jb{Note that the resulting survey} sample
was more than ten times larger than that of a recent attempt to \jb{characterize} the features of the top 10 most cited scientific publications authored by biomedical researchers~\cite{ioannidis2014your}.

\section{Results}

\begin{figure}
\begin{center}
\includegraphics[width = 0.45\textwidth]{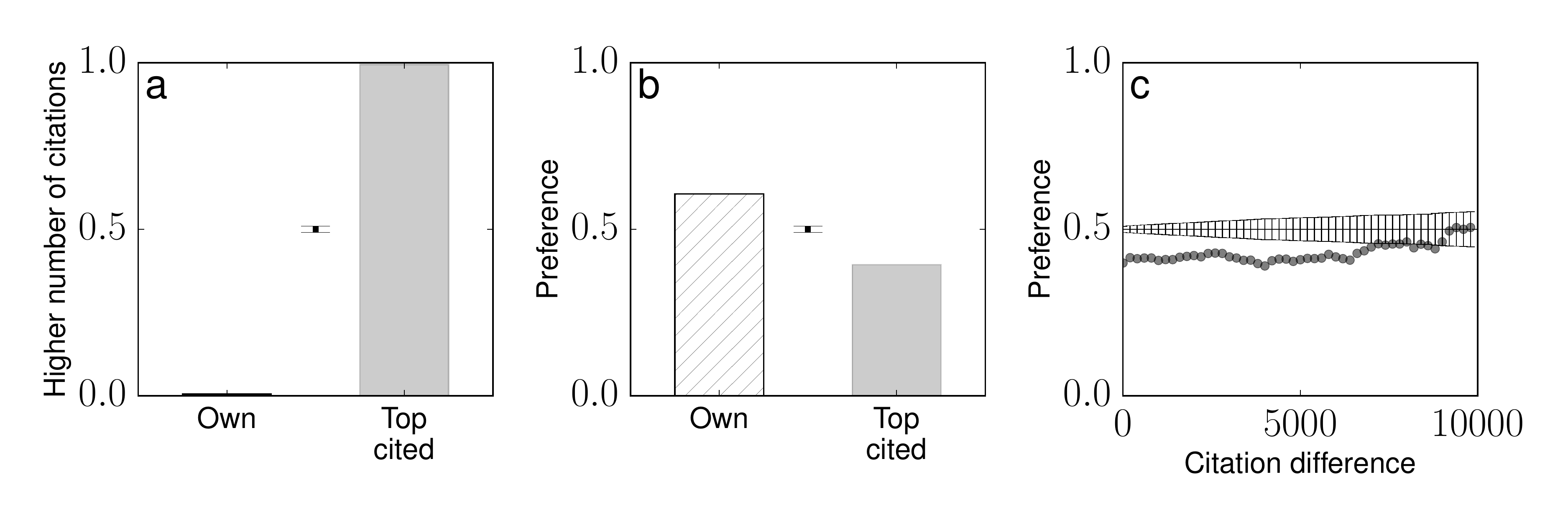}
\end{center}
\caption{
\noindent {\bf Perceived impact of authored {\it vs.} top-cited papers.}
We consider $N = 2,916$ comparisons between
one paper taken from 
the ``Own'' pool, and the other one extracted 
from the ``Top cited'' pool.
{\bf a} Probability $P(c_a > c_b | p_a=o, p_b=t)$ 
that the paper written by the respondent
accumulated more citations than the top-cited article.
To estimate such a probability, we consider only
comparisons between papers with different 
number of citations, so that the 
difference in citations is always different from zero.
We obtain $P(c_a > c_b | p_a=o, p_b=t) = 0.005$, corresponding to
a standard score $|z| = 53.36$ with respect 
to the unbiased binomial model.
The error bar in the graph is centered at the unbiased expected 
value $0.5$, and has height equal to twice the value of the 
standard deviation $\sigma = 0.009$.
{\bf b} Probability $P(a \to b | p_a=o, p_b=t)$
that the respondent preferred 
her/his own paper to the one taken from the
pool of Top cited articles.
Such a probability amounts to
$P(a \to b | p_a=o, p_b=t) = 0.61$, corresponding to $|z| = 11.48$. 
{\bf c}
Probability $P_{ot}(a \to b | c_a > c_b, c_a-c_b \geq \Delta c)$
that the respondent preferred the paper with more citations. 
Each point represents
the probability of preference for the
paper with higher number of citations for
all pairs of papers whose 
citation difference $\Delta c$ is higher
than the value reported on the x-axis.
The horizontal line indicates the naive expectation $0.5$
of the unbiased binomial model.
Height of the error bars equals two standard deviations of the
binomial model.
}
\label{fig1}
\end{figure}

Figure~\ref{fig1} summarizes the relation between
citation and perceived impact obtained
from the analysis of comparisons between one
paper taken from the OWN and 
the other from the TCD pool.
Naturally, a paper in the TCD
pool is very likely to have a citation impact
larger than an article of the OWN, as the measure of 
the probability $P(c_a > c_b | p_a=o, p_b=t)$
shows in Fig.~\ref{fig1}a. Here, $c_x$ denotes the total 
number of citations accumulated
by paper $x$, and $p_x$ denotes the pool of the paper 
$x$ ($o$ stands for OWN, and $t$ for TCD).
Nonetheless, perceived impact of articles written by respondents 
themselves is generally higher than the ones of papers from the TCD pool, revealing a strong bias towards respondents\jb{'} own papers.
This is visualized in Fig.~\ref{fig1}B, where we
consider the probability $P(a \to b | p_a = o, p_b = t)$, i.e., the probability
that an OWN article was preferred to one from the TCD pool
(the notation $x \to y$ denotes preference of paper $x$ with 
respect to paper $y$).
To quantify the statistical significance of our measurements, 
we use absolute values of the standard score 
$z$ with respect to an unbiased binomial distribution, 
hence $z = (P - 0.5) / \sigma$, where $P$ is the actual
value of the probability measured in the experiment,
$\sigma = \sqrt{0.5\, \times \, 0.5/N}$, and
$N$ is the size of the sample (i.e., number of comparisons).
Note that, since only two possibilities are available,
it doesn't matter if we measure the
probability $P$ or its complementary probability 
$1-P$. The absolute value of the standard 
score $|z|$ doesn't depend on this choice.
The statistical interpretation of the standard score
for an unbiased binomial distribution with sufficiently
large sample sizes, as in our case, 
is similar to the one valid for the standard normal 
distribution, so that $p$-values $< 0.001$ 
approximately correspond to $|z| > 3$, and the $p$-value
decreases exponentially fast as $|z|$ increases.
The empirical results of Fig.~\ref{fig1}B are highly unlikely 
to happen by chance. In fact, we have
$P(a \to b | p_a = o, p_b = t) = 0.61$ and $N = 2,916$, 
leading to $|z| = 11.48$.  One  may wonder whether this result is 
actually dependent on the difference in
citation impact between the two papers 
or not. Fig.~\ref{fig1}C points to a possible answer to this question.
We consider the probability $P_{ot}(a \to b | c_a > c_b, c_a - c_b \geq \Delta c)$ that respondents preferred
the more cited paper among the two articles in the comparison 
as a function of the difference $\Delta c$ in citations between the two papers
(for \jb{brevity} we used the suffix $ot$ to indicate
the pools where the two papers were taken \jb{from}).
Surprisingly, citation and perceived impact are negatively
correlated, in a statistically significant manner, for a wide range
of values of the difference of citations accumulated by the
two papers. Only when the citation impact of the TCD 
paper is much larger than the one 
of the OWN paper, we do not longer observe a statistically significant preference. 
The fact that scholars tended
to systematically prefer their own articles regardless of the 
comparison can be interpreted as the consequence of an 
egocentric~\cite{ross1979egocentric} or familiarity bias \cite{zajonc1968attitudinal}.
Egocentric bias doesn't necessarily have
a negative connotation. The bias could be reconcilable 
with the fact that researchers base most of their work on results of 
their own past research, and they might have interpreted the generic question posed in the survey in this way. Furthermore, since researchers
are inherently most familiar with their own work, the uncertainty with regards to their impact might be lower and hence these articles might enjoy the authors' preference over a paper that is less familiar and whose impact is thus more difficult to assess reliably.

\begin{figure}
\begin{center}
\includegraphics[width = 0.45\textwidth]{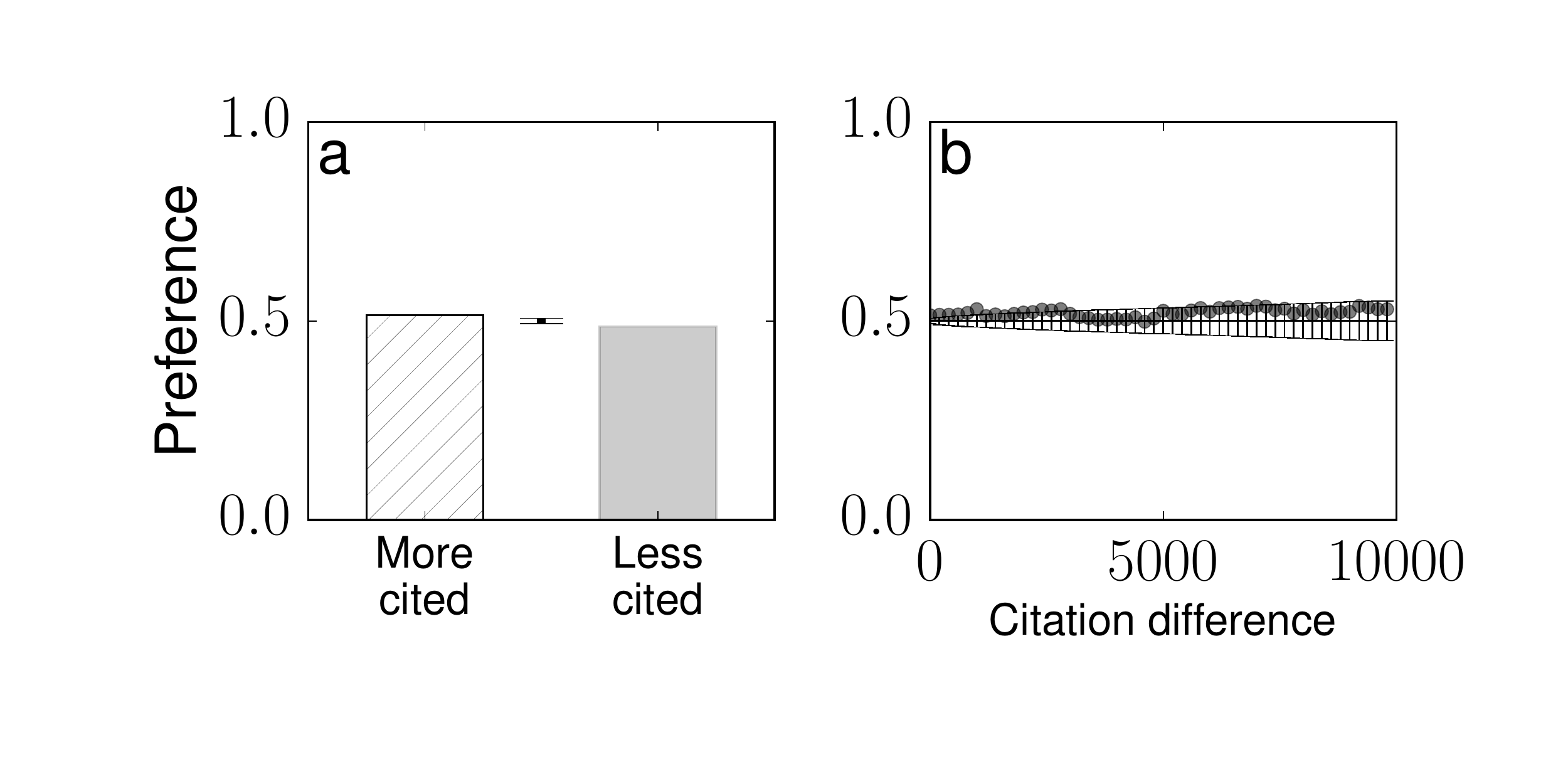}
\end{center}
\caption{
\noindent {\bf Perceived impact of top-cited
papers.}
We consider only comparisons between
two papers taken from the ``Top cited'' pool.
{\bf a} Probability $P_{tt}(a \to b| c_a > c_b)$ that the respondent preferred
the paper with higher/lower number of citations. 
Papers with higher citation counts are preferred with
probability  $P_{tt}(a \to b| c_a > c_b) = 0.51$, calculated over a sample 
of size $N = 5,930$, leading to a standard score
$|z| = 2.18$. Error bar is centered at $0.5$ and has height
equal to twice the standard deviation $\sigma$ of the 
unbiased binomial model. Here $\sigma = 0.007$. 
{\bf b}
Probability $P_{tt}(a \to b| c_a > c_b, c_a-c_b \geq \Delta c)$ that the respondent preferred
the paper with more citations as a function
of the difference in citations among the papers.
The horizontal line indicates the naive expectation of $0.5$
of the unbiased binomial model.
Height of the error bars equals two standard deviations of the
binomial model.
}
\label{fig2}
\end{figure}

To discount for the presence of an egocentric or familiarity bias, we 
turn our attention to results obtained from comparisons where 
both papers were extracted
from the same pool. We start from the description
of our findings regarding the TCD pool (Fig.~\ref{fig2}).
Here, a positive correlation is observed
between citation impact and perceived impact.
Overall, papers with higher citation impact
were preferred at higher rates by researchers,
but only with probability $P_{tt}(a \to b | c_a > c_b)=0.51$ 
leading to a standard score $|z|=2.18$ ($p$-value = $0.02$, 
Fig.~\ref{fig2}a). Also,
perceived impact doesn't vary significantly
with citation impact even if the difference
in citations received by papers in the comparison
can be larger than three orders of magnitude (Fig.~\ref{fig2}b). 
\filippo{The result seems not dependent on the age
of papers, as we do not observe any systematic preference (Fig.~\ref{figSM3A}).}
We speculate that, from the perspective of individual researchers, citing a top-cited article in their own research area may be 
interpreted  as a public ``homage'' to the collective popularity of 
the paper, and thus reconcilable with a cumulative advantage principle~\cite{price1976general, barabasi1999emergence, wang2013quantifying}, the conformity bias \cite{asch1956}, or
a copying behaviour~\cite{krapivsky2005network, simkin2002read}, rather than a true recognition of the importance of 
the work for their own research.

\begin{figure}
\begin{center}
\includegraphics[width = 0.45\textwidth]{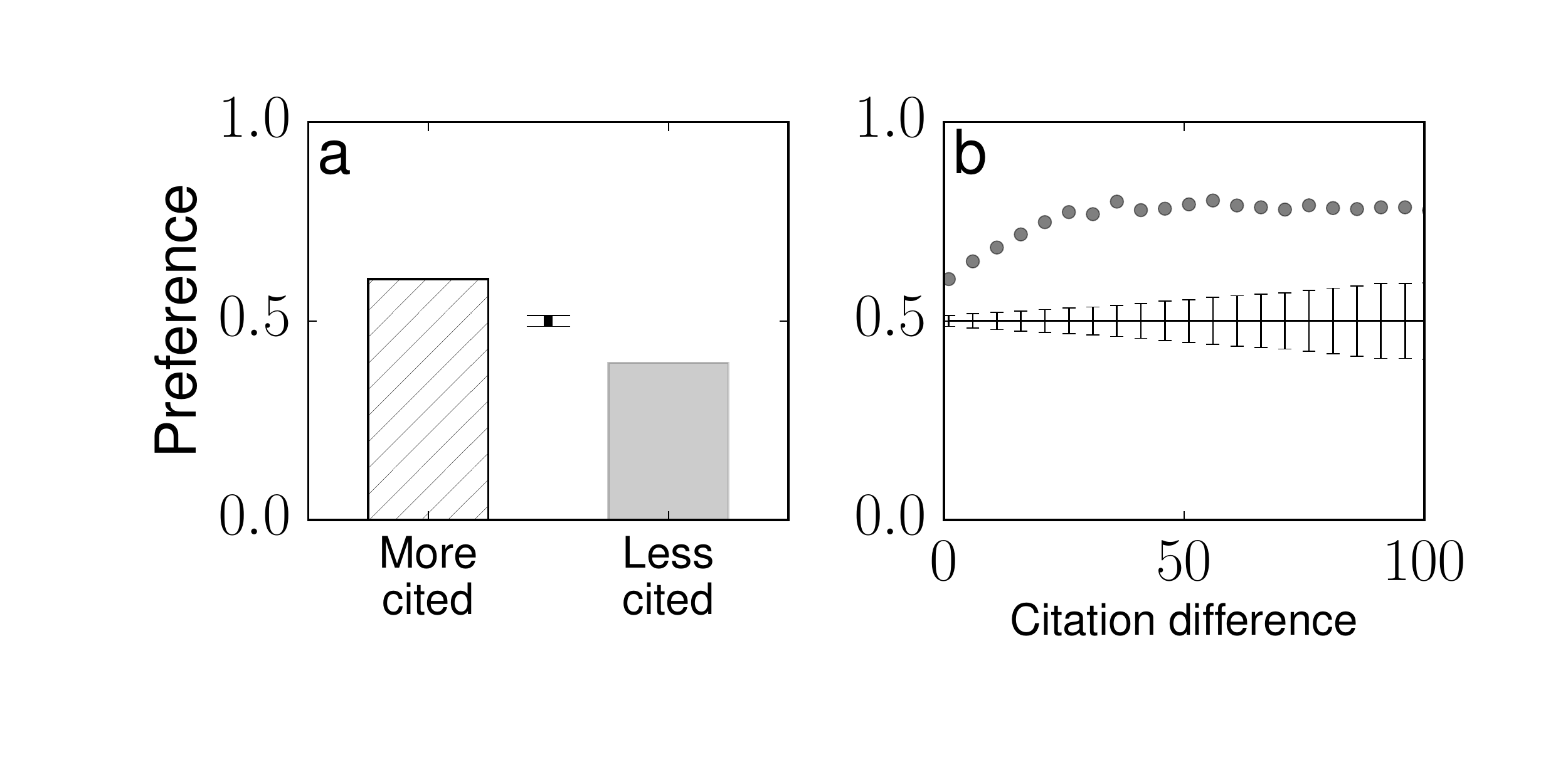}
\end{center}
\caption{
\noindent {\bf Perceived impact of authored papers.}
We consider only comparisons between
two papers taken from the ``Own'' pool.
{\bf A} Probability $P_{oo}(a \to b| c_a > c_b)$ that the respondent preferred
the paper with higher/lower number of citations. 
Papers with higher citation counts are preferred with
probability $P_{oo}(a \to b| c_a > c_b) = 0.61$
calculated over a sample 
of size $N = 1,302$, leading to a standard score
$|z| =  7.60$. Error bar is centered at $0.5$ and has height
equal to twice the standard deviation $\sigma$ of the 
unbiased binomial model. Here $\sigma = 0.014$. 
{\bf B}
Probability $P_{oo}(a \to b| c_a > c_b, c_a-c_b \geq \Delta c)$ that the respondent preferred
the paper with more citations as a function
of the difference in citations among the papers. 
The horizontal line indicates the naive expectation of $0.5$
of the unbiased binomial model.
Height of the error bars equals two standard deviations of the
binomial model.
}
\label{fig3}
\end{figure}
Figs.~\ref{fig1} and~\ref{fig2}
suggest that the total number of citations received 
by a paper doesn't reliably quantify the true impact or influence
that the paper has for the actual research of a scholar. 
Such a conclusion is, however, overturned when
we consider pairwise comparisons between 
articles from the OWN pool (Fig.~\ref{fig3}).
In this case, we observe a significant alignment
between citation impact and expert impact as researchers
systematically preferred the highest cited articles among their own articles.
Preference to the more cited article was given
with probability $P_{oo}(a \to b | c_a > c_b)=0.61$, corresponding to
$|z|=7.60$ (Fig.~\ref{fig3}a). 
Furthermore, we observed a systematic increase 
in the preference for the more cited paper as a function
of the difference of citation impact between the papers
in the comparison (Fig.~\ref{fig3}b). Preference values
saturate to almost $0.75$, if citations accumulated 
by papers differ by $50$ or more. This effect is
even stronger when one accounts for the general tendency
of preference for more recent publications, and
the fact that more recent publications generally exhibit lower
citation impact (Fig.~\ref{figSM3}). Overall, this finding
conclusively supports the observations by Ioannidis {\it et al.} in
their small-scale survey~\cite{ioannidis2014your}. 

\begin{figure}
\begin{center}
\includegraphics[width = 0.45\textwidth]{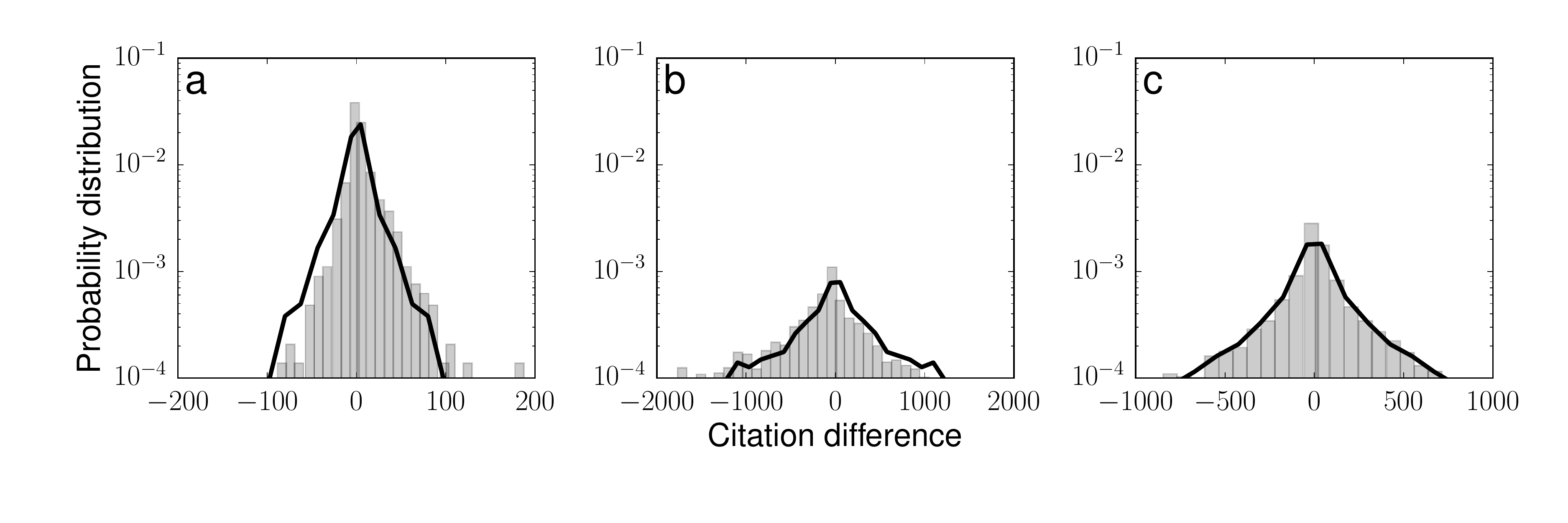}
\end{center}
\caption{
\noindent {\bf Citation impact of articles with higher perceived impact.}
{\bf a} We consider only comparisons between
two papers taken from the ``Own'' pool as in 
Fig.~\ref{fig3}. We include here also comparisons between
papers having the same number of citations, so that the
difference of their citation impacts can be equal to zero.
The gray bars stand for the probability distribution $P_{oo} (c_a - c_b = \Delta c | a \to b)$ 
of the difference in citation
impact between preferred and non-preferred article in 
the comparison. The average value of the 
distribution is $11.16$, the
standard deviation is $130.50$, and the skewness is $29.28$.
The black line serves as a term of comparison as it represents
the distribution $P_{oo} (c_a - c_b = \Delta c)$ of difference in citation impact between
pairs of papers in the comparisons, irrespective of the preference
expressed by researchers. This distribution is by definition
symmetric (null skewness), and centered in 
zero (average value equal to zero).
Probability distributions are normalized such that the
integral below the curves equals one.
{\bf b}
Same as in panel a but for comparisons
between one paper taken from the ``Own'' pool
and the other extracted from the ``Top cited'' pool.
Data are the same as those used in Fig.~\ref{fig1}.
The average value of the distribution $P_{ot} (c_a - c_b = \Delta c | a \to b)$  is
$-267.50$, the standard deviation is $4762.69$, and 
the skewness is $-1.36$. The black line stands for the
unconditional probability $P_{ot} (c_a - c_b = \Delta c)$.
{\bf c} Same as in panels A and B but for comparisons
between two papers taken from the ``Top cited'' pool.
Data are the same as those used in Fig.~\ref{fig2}.
The average value of the distribution $P_{tt} (c_a - c_b = \Delta c | a \to b)$  is
$99.47$, the standard deviation is $4255.18$, 
and the skewness is $12.34$. The black line stands for the
unconditional probability $P_{tt} (c_a - c_b = \Delta c)$.
}
\label{fig4}
\end{figure}

Our observations are further supported by the
results presented in Fig.~\ref{fig4}, where we consider
the distribution of the difference in citation impact
between preferred and non-preferred articles in 
different types of comparisons. The distribution
is clearly right-skewed for comparisons between papers
in the pool of OWN articles (Fig.~\ref{fig4}a). 
The distribution for comparisons between articles from the
OWN and TCD pools on the other hand has negative average value,
and is negatively skewed (Fig.~\ref{fig4}b). Finally, the distribution
in Fig.~\ref{fig4}c,
for comparisons between papers in the TCD pool,
provides evidence of a slight preference for articles with higher 
number of citations, but the preference is
much less evident than in the case of Fig.~\ref{fig4}a.

\section{Conclusions}

\filippo{
In summary, our work  quantifies the degree of correlation between citation impact and a new 
post-publication metric, namely impact as perceived by the authors themselves. 
Our survey serves to understand whether citation numbers ``truly'' reflect the impact of papers as it pertains to the 
daily practice of researchers. This is a particular
dimension of impact that has been not quantified \jb{before}.
}
The results from our large-scale
survey generate one very important conclusion:
citation numbers
approximate with good accuracy the \filippo{perceived}
impact of scientific publications, but only when 
psychological biases that make authors prefer their own 
papers above those of others are removed. 
This conclusion can be justified by
the reasonable assumption 
that the self-assessment of one's own papers 
is the most objective (and least
biased) quantification 
of perceived impact since authors are assumed to know 
their own articles
best and are thus the best 
judges of their comparative impact. Another possible explanation 
could be the Goodhart effect~\cite{goodhart1984problems}, namely 
that the preferences of authors are shaped by the knowledge 
of citations accumulated by their own papers. As a result, pairwise 
assessments of their own papers will be congruent with citation data.

\section*{Acknowledgements}
We are grateful to all researchers who took part in our survey.
This work uses Web of Science data by Thomson Reuters, provided by the 
Network Science Institute at Indiana University. 
This work is funded by  the National Science Foundation (grant SMA-1446078 and SMA-1636636).


\begin{thebibliography}{32}
\expandafter\ifx\csname natexlab\endcsname\relax\def\natexlab#1{#1}\fi
\expandafter\ifx\csname bibnamefont\endcsname\relax
  \def\bibnamefont#1{#1}\fi
\expandafter\ifx\csname bibfnamefont\endcsname\relax
  \def\bibfnamefont#1{#1}\fi
\expandafter\ifx\csname citenamefont\endcsname\relax
  \def\citenamefont#1{#1}\fi
\expandafter\ifx\csname url\endcsname\relax
  \def\url#1{\texttt{#1}}\fi
\expandafter\ifx\csname urlprefix\endcsname\relax\def\urlprefix{URL }\fi
\providecommand{\bibinfo}[2]{#2}
\providecommand{\eprint}[2][]{\url{#2}}

\bibitem[{\citenamefont{Harzing}(2010)}]{harzing2010publish}
\bibinfo{author}{\bibfnamefont{A.-W.} \bibnamefont{Harzing}},
  \emph{\bibinfo{title}{The publish or perish book}} (\bibinfo{publisher}{Tarma
  Software Research}, \bibinfo{year}{2010}).

\bibitem[{\citenamefont{Bornmann and Daniel}(2006)}]{bor2006}
\bibinfo{author}{\bibfnamefont{L.}~\bibnamefont{Bornmann}} \bibnamefont{and}
  \bibinfo{author}{\bibfnamefont{H.-D.} \bibnamefont{Daniel}},
  \bibinfo{journal}{Scientometrics} \textbf{\bibinfo{volume}{68}},
  \bibinfo{pages}{427} (\bibinfo{year}{2006}), ISSN \bibinfo{issn}{0138-9130}.

\bibitem[{\citenamefont{Bornmann et~al.}(2008)\citenamefont{Bornmann, Wallon,
  and Ledin}}]{bornmann2008does}
\bibinfo{author}{\bibfnamefont{L.}~\bibnamefont{Bornmann}},
  \bibinfo{author}{\bibfnamefont{G.}~\bibnamefont{Wallon}}, \bibnamefont{and}
  \bibinfo{author}{\bibfnamefont{A.}~\bibnamefont{Ledin}},
  \bibinfo{journal}{PLoS One} \textbf{\bibinfo{volume}{3}},
  \bibinfo{pages}{e3480} (\bibinfo{year}{2008}).

\bibitem[{\citenamefont{Lovegrove and Johnson}(2008)}]{lovegrove2008assessment}
\bibinfo{author}{\bibfnamefont{B.~G.} \bibnamefont{Lovegrove}}
  \bibnamefont{and} \bibinfo{author}{\bibfnamefont{S.~D.}
  \bibnamefont{Johnson}}, \bibinfo{journal}{Bioscience}
  \textbf{\bibinfo{volume}{58}}, \bibinfo{pages}{160} (\bibinfo{year}{2008}).

\bibitem[{\citenamefont{Hornbostel et~al.}(2009)\citenamefont{Hornbostel,
  B{\"o}hmer, Klingsporn, Neufeld, and von Ins}}]{hornbostel2009funding}
\bibinfo{author}{\bibfnamefont{S.}~\bibnamefont{Hornbostel}},
  \bibinfo{author}{\bibfnamefont{S.}~\bibnamefont{B{\"o}hmer}},
  \bibinfo{author}{\bibfnamefont{B.}~\bibnamefont{Klingsporn}},
  \bibinfo{author}{\bibfnamefont{J.}~\bibnamefont{Neufeld}}, \bibnamefont{and}
  \bibinfo{author}{\bibfnamefont{M.}~\bibnamefont{von Ins}},
  \bibinfo{journal}{Scientometrics} \textbf{\bibinfo{volume}{79}},
  \bibinfo{pages}{171} (\bibinfo{year}{2009}).

\bibitem[{\citenamefont{Bornmann et~al.}(2011)\citenamefont{Bornmann, Mutz,
  Marx, Schier, and Daniel}}]{bornmann2011multilevel}
\bibinfo{author}{\bibfnamefont{L.}~\bibnamefont{Bornmann}},
  \bibinfo{author}{\bibfnamefont{R.}~\bibnamefont{Mutz}},
  \bibinfo{author}{\bibfnamefont{W.}~\bibnamefont{Marx}},
  \bibinfo{author}{\bibfnamefont{H.}~\bibnamefont{Schier}}, \bibnamefont{and}
  \bibinfo{author}{\bibfnamefont{H.-D.} \bibnamefont{Daniel}},
  \bibinfo{journal}{Journal of the Royal Statistical Society: Series A
  (Statistics in Society)} \textbf{\bibinfo{volume}{174}}, \bibinfo{pages}{857}
  (\bibinfo{year}{2011}).

\bibitem[{\citenamefont{Bar-Ilan}(2008)}]{bar2008informetrics}
\bibinfo{author}{\bibfnamefont{J.}~\bibnamefont{Bar-Ilan}},
  \bibinfo{journal}{Journal of Informetrics} \textbf{\bibinfo{volume}{2}},
  \bibinfo{pages}{1} (\bibinfo{year}{2008}).

\bibitem[{\citenamefont{Van~Noorden}(2010)}]{van2010metrics}
\bibinfo{author}{\bibfnamefont{R.}~\bibnamefont{Van~Noorden}},
  \bibinfo{journal}{Nature} \textbf{\bibinfo{volume}{465}},
  \bibinfo{pages}{864} (\bibinfo{year}{2010}).

\bibitem[{\citenamefont{Hirsch}(2005)}]{hirsch2005index}
\bibinfo{author}{\bibfnamefont{J.~E.} \bibnamefont{Hirsch}},
  \bibinfo{journal}{Proceedings of the National academy of Sciences of the
  United States of America} \textbf{\bibinfo{volume}{102}},
  \bibinfo{pages}{16569} (\bibinfo{year}{2005}).

\bibitem[{\citenamefont{Garfield}(2006)}]{garfield2006history}
\bibinfo{author}{\bibfnamefont{E.}~\bibnamefont{Garfield}},
  \bibinfo{journal}{JAMA: the journal of the American Medical Association}
  \textbf{\bibinfo{volume}{295}}, \bibinfo{pages}{90} (\bibinfo{year}{2006}).

\bibitem[{\citenamefont{Bornmann and
  Daniel}(2008{\natexlab{a}})}]{bornmann2008citation}
\bibinfo{author}{\bibfnamefont{L.}~\bibnamefont{Bornmann}} \bibnamefont{and}
  \bibinfo{author}{\bibfnamefont{H.-D.} \bibnamefont{Daniel}},
  \bibinfo{journal}{Journal of Documentation} \textbf{\bibinfo{volume}{64}},
  \bibinfo{pages}{45} (\bibinfo{year}{2008}{\natexlab{a}}).

\bibitem[{\citenamefont{Sarig{\"o}l et~al.}(2014)\citenamefont{Sarig{\"o}l,
  Pfitzner, Scholtes, Garas, and Schweitzer}}]{sarigol2014predicting}
\bibinfo{author}{\bibfnamefont{E.}~\bibnamefont{Sarig{\"o}l}},
  \bibinfo{author}{\bibfnamefont{R.}~\bibnamefont{Pfitzner}},
  \bibinfo{author}{\bibfnamefont{I.}~\bibnamefont{Scholtes}},
  \bibinfo{author}{\bibfnamefont{A.}~\bibnamefont{Garas}}, \bibnamefont{and}
  \bibinfo{author}{\bibfnamefont{F.}~\bibnamefont{Schweitzer}},
  \bibinfo{journal}{arXiv preprint arXiv:1402.7268}  (\bibinfo{year}{2014}).

\bibitem[{\citenamefont{Garfield}(1979)}]{Garfield1979}
\bibinfo{author}{\bibfnamefont{E.}~\bibnamefont{Garfield}},
  \bibinfo{journal}{Scientometrics} \textbf{\bibinfo{volume}{1}},
  \bibinfo{pages}{359} (\bibinfo{year}{1979}).

\bibitem[{\citenamefont{MacRoberts and
  MacRoberts}(1996)}]{macroberts1996problems}
\bibinfo{author}{\bibfnamefont{M.~H.} \bibnamefont{MacRoberts}}
  \bibnamefont{and} \bibinfo{author}{\bibfnamefont{B.~R.}
  \bibnamefont{MacRoberts}}, \bibinfo{journal}{Scientometrics}
  \textbf{\bibinfo{volume}{36}}, \bibinfo{pages}{435} (\bibinfo{year}{1996}).

\bibitem[{\citenamefont{Adler et~al.}(2009)\citenamefont{Adler, Ewing, and
  Taylor}}]{adler2009}
\bibinfo{author}{\bibfnamefont{R.}~\bibnamefont{Adler}},
  \bibinfo{author}{\bibfnamefont{J.}~\bibnamefont{Ewing}}, \bibnamefont{and}
  \bibinfo{author}{\bibfnamefont{P.}~\bibnamefont{Taylor}},
  \bibinfo{journal}{Statistical Science} \textbf{\bibinfo{volume}{24}},
  \bibinfo{pages}{1} (\bibinfo{year}{2009}).

\bibitem[{\citenamefont{MacRoberts and
  MacRoberts}(2010)}]{macroberts2010problems}
\bibinfo{author}{\bibfnamefont{M.~H.} \bibnamefont{MacRoberts}}
  \bibnamefont{and} \bibinfo{author}{\bibfnamefont{B.~R.}
  \bibnamefont{MacRoberts}}, \bibinfo{journal}{Journal of the American Society
  for Information Science and Technology} \textbf{\bibinfo{volume}{61}},
  \bibinfo{pages}{1} (\bibinfo{year}{2010}).

\bibitem[{\citenamefont{Wuchty et~al.}(2007)\citenamefont{Wuchty, Jones, and
  Uzzi}}]{wuchty2007increasing}
\bibinfo{author}{\bibfnamefont{S.}~\bibnamefont{Wuchty}},
  \bibinfo{author}{\bibfnamefont{B.~F.} \bibnamefont{Jones}}, \bibnamefont{and}
  \bibinfo{author}{\bibfnamefont{B.}~\bibnamefont{Uzzi}},
  \bibinfo{journal}{Science} \textbf{\bibinfo{volume}{316}},
  \bibinfo{pages}{1036} (\bibinfo{year}{2007}).

\bibitem[{\citenamefont{Radicchi et~al.}(2008)\citenamefont{Radicchi,
  Fortunato, and Castellano}}]{radicchi2008universality}
\bibinfo{author}{\bibfnamefont{F.}~\bibnamefont{Radicchi}},
  \bibinfo{author}{\bibfnamefont{S.}~\bibnamefont{Fortunato}},
  \bibnamefont{and}
  \bibinfo{author}{\bibfnamefont{C.}~\bibnamefont{Castellano}},
  \bibinfo{journal}{Proceedings of the National Academy of Sciences USA}
  \textbf{\bibinfo{volume}{105}}, \bibinfo{pages}{17268}
  (\bibinfo{year}{2008}).

\bibitem[{\citenamefont{Wang et~al.}(2013)\citenamefont{Wang, Song, and
  Barab{\'a}si}}]{wang2013quantifying}
\bibinfo{author}{\bibfnamefont{D.}~\bibnamefont{Wang}},
  \bibinfo{author}{\bibfnamefont{C.}~\bibnamefont{Song}}, \bibnamefont{and}
  \bibinfo{author}{\bibfnamefont{A.-L.} \bibnamefont{Barab{\'a}si}},
  \bibinfo{journal}{Science} \textbf{\bibinfo{volume}{342}},
  \bibinfo{pages}{127} (\bibinfo{year}{2013}).

\bibitem[{\citenamefont{Bornmann and
  Daniel}(2008{\natexlab{b}})}]{citations:bornman2008}
\bibinfo{author}{\bibfnamefont{L.}~\bibnamefont{Bornmann}} \bibnamefont{and}
  \bibinfo{author}{\bibfnamefont{H.}~\bibnamefont{Daniel}},
  \bibinfo{journal}{Journal of Documentation} \textbf{\bibinfo{volume}{6}},
  \bibinfo{pages}{45} (\bibinfo{year}{2008}{\natexlab{b}}).

\bibitem[{\citenamefont{Bollen et~al.}(2009)\citenamefont{Bollen, Van~de
  Sompel, Hagberg, and Chute}}]{bollen2009principal}
\bibinfo{author}{\bibfnamefont{J.}~\bibnamefont{Bollen}},
  \bibinfo{author}{\bibfnamefont{H.}~\bibnamefont{Van~de Sompel}},
  \bibinfo{author}{\bibfnamefont{A.}~\bibnamefont{Hagberg}}, \bibnamefont{and}
  \bibinfo{author}{\bibfnamefont{R.}~\bibnamefont{Chute}},
  \bibinfo{journal}{PloS one} \textbf{\bibinfo{volume}{4}},
  \bibinfo{pages}{e6022} (\bibinfo{year}{2009}).

\bibitem[{\citenamefont{Sheehan}(2001)}]{JCC4:JCC403}
\bibinfo{author}{\bibfnamefont{K.~B.} \bibnamefont{Sheehan}},
  \bibinfo{journal}{Journal of Computer-Mediated Communication}
  \textbf{\bibinfo{volume}{6}}, \bibinfo{pages}{0} (\bibinfo{year}{2001}).

\bibitem[{\citenamefont{Bethlehem}(2010)}]{bethlehem2010selection}
\bibinfo{author}{\bibfnamefont{J.}~\bibnamefont{Bethlehem}},
  \bibinfo{journal}{International Statistical Review}
  \textbf{\bibinfo{volume}{78}}, \bibinfo{pages}{161} (\bibinfo{year}{2010}).

\bibitem[{\citenamefont{Ioannidis et~al.}(2014)\citenamefont{Ioannidis, Boyack,
  Small, Sorensen, and Klavans}}]{ioannidis2014your}
\bibinfo{author}{\bibfnamefont{J.}~\bibnamefont{Ioannidis}},
  \bibinfo{author}{\bibfnamefont{K.~W.} \bibnamefont{Boyack}},
  \bibinfo{author}{\bibfnamefont{H.}~\bibnamefont{Small}},
  \bibinfo{author}{\bibfnamefont{A.~A.} \bibnamefont{Sorensen}},
  \bibnamefont{and} \bibinfo{author}{\bibfnamefont{R.}~\bibnamefont{Klavans}},
  \bibinfo{journal}{Nature} \textbf{\bibinfo{volume}{514}},
  \bibinfo{pages}{561} (\bibinfo{year}{2014}).

\bibitem[{\citenamefont{Ross and Sicoly}(1979)}]{ross1979egocentric}
\bibinfo{author}{\bibfnamefont{M.}~\bibnamefont{Ross}} \bibnamefont{and}
  \bibinfo{author}{\bibfnamefont{F.}~\bibnamefont{Sicoly}},
  \bibinfo{journal}{Journal of personality and social psychology}
  \textbf{\bibinfo{volume}{37}}, \bibinfo{pages}{322} (\bibinfo{year}{1979}).

\bibitem[{\citenamefont{Zajonc}(1968)}]{zajonc1968attitudinal}
\bibinfo{author}{\bibfnamefont{R.~B.} \bibnamefont{Zajonc}},
  \bibinfo{journal}{Journal of personality and social psychology}
  \textbf{\bibinfo{volume}{9}}, \bibinfo{pages}{1} (\bibinfo{year}{1968}).

\bibitem[{\citenamefont{Price}(1976)}]{price1976general}
\bibinfo{author}{\bibfnamefont{D.~d.~S.} \bibnamefont{Price}},
  \bibinfo{journal}{Journal of the American society for Information science}
  \textbf{\bibinfo{volume}{27}}, \bibinfo{pages}{292} (\bibinfo{year}{1976}).

\bibitem[{\citenamefont{Barab{\'a}si and Albert}(1999)}]{barabasi1999emergence}
\bibinfo{author}{\bibfnamefont{A.-L.} \bibnamefont{Barab{\'a}si}}
  \bibnamefont{and} \bibinfo{author}{\bibfnamefont{R.}~\bibnamefont{Albert}},
  \bibinfo{journal}{science} \textbf{\bibinfo{volume}{286}},
  \bibinfo{pages}{509} (\bibinfo{year}{1999}).

\bibitem[{\citenamefont{Asch}(1956)}]{asch1956}
\bibinfo{author}{\bibfnamefont{S.~E.} \bibnamefont{Asch}},
  \bibinfo{journal}{Psychological Monographs} \textbf{\bibinfo{volume}{70}},
  \bibinfo{pages}{1} (\bibinfo{year}{1956}).

\bibitem[{\citenamefont{Krapivsky and Redner}(2005)}]{krapivsky2005network}
\bibinfo{author}{\bibfnamefont{P.}~\bibnamefont{Krapivsky}} \bibnamefont{and}
  \bibinfo{author}{\bibfnamefont{S.}~\bibnamefont{Redner}},
  \bibinfo{journal}{Physical Review E} \textbf{\bibinfo{volume}{71}},
  \bibinfo{pages}{036118} (\bibinfo{year}{2005}).

\bibitem[{\citenamefont{Simkin and Roychowdhury}(2002)}]{simkin2002read}
\bibinfo{author}{\bibfnamefont{M.~V.} \bibnamefont{Simkin}} \bibnamefont{and}
  \bibinfo{author}{\bibfnamefont{V.~P.} \bibnamefont{Roychowdhury}},
  \bibinfo{journal}{arXiv preprint cond-mat/0212043}  (\bibinfo{year}{2002}).

\bibitem[{\citenamefont{Goodhart}(1984)}]{goodhart1984problems}
\bibinfo{author}{\bibfnamefont{C.~A.} \bibnamefont{Goodhart}},
  \emph{\bibinfo{title}{Problems of monetary management: the UK experience}}
  (\bibinfo{publisher}{Springer}, \bibinfo{year}{1984}).

\end{thebibliography}

\clearpage


\newpage

\renewcommand{\theequation}{A\arabic{equation}}
\setcounter{equation}{0}
\renewcommand{\thefigure}{A\arabic{figure}}
\setcounter{figure}{0}
\renewcommand{\thetable}{A\arabic{table}}
\setcounter{table}{0}

\setcounter{page}{1}

\section*{Appendix}

\onecolumngrid

\begin{figure}[!h]
\begin{center}
\includegraphics[width = 0.7\textwidth]{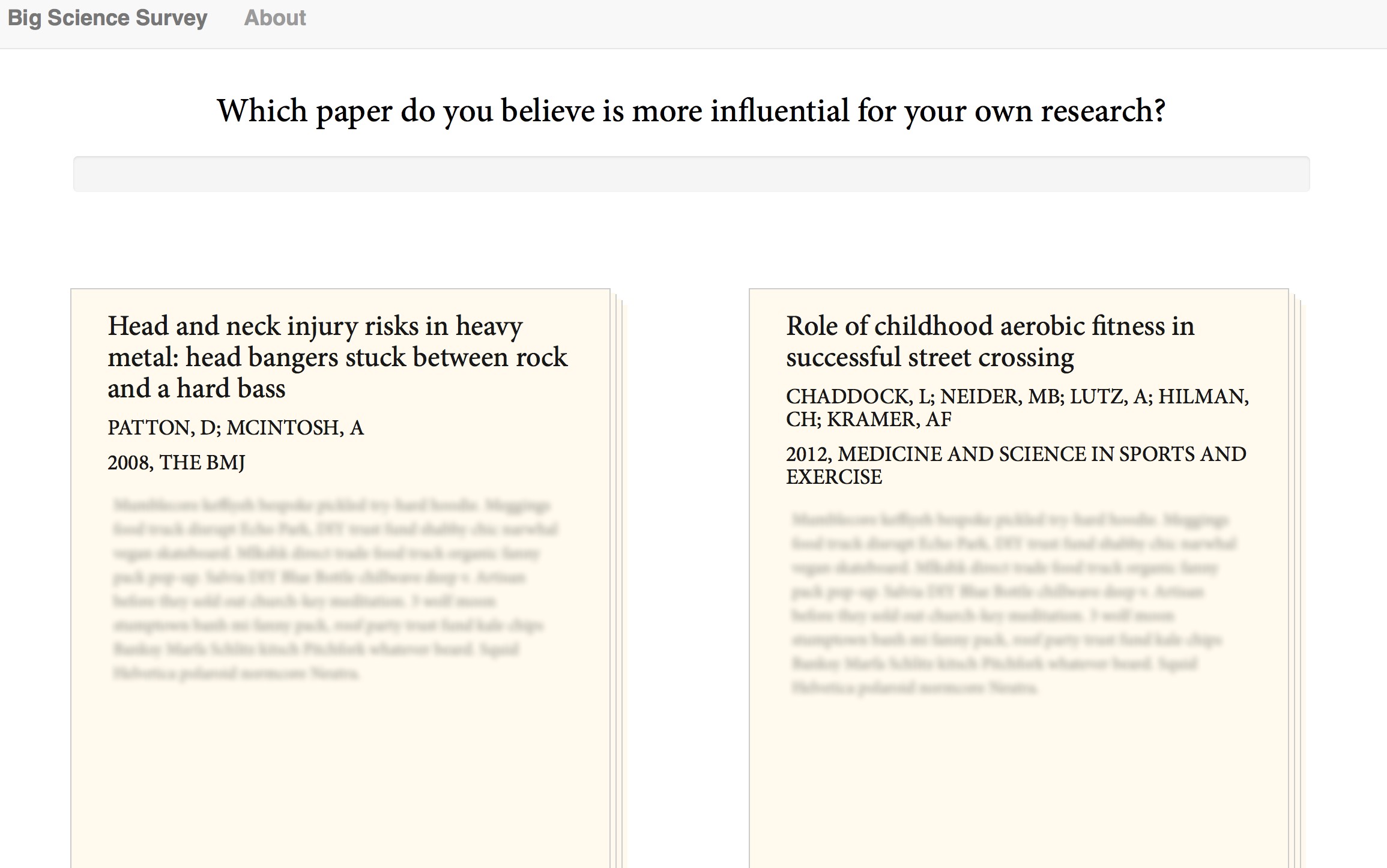}
\end{center}
\caption{
\noindent {\bf Survey design.}
Screenshot from the online survey at \url{bigscience.soic.indiana.edu}.
The visualization shows an example of a comparison presented to
individual participants. Papers visualized in this
example were randomly chosen from our dataset.
}
\label{figSURVEY}
\end{figure}

\begin{figure}
    \begin{mdframed}[style=mystyle]
            \begin{center}
	            \begin{verbatim}
Hello,

You are receiving this email because you are the 
corresponding author of several scientific publications. 
We are conducting a survey, the largest of its kind, 
to determine the relative importance of scientific 
publications as perceived by experienced researchers.

We've generated a survey specifically for you that 
consists of anywhere between 1 and 10 pairs of papers 
drawn from our database. The publications may be drawn 
from your own publications or publications you've cited.  
For each pair, we ask you to select the publication that 
you believe is more influential for your own research.

It should take less than two minutes to complete the 
ten comparisons. To participate, please follow this link:

bigscience.soic.indiana.edu/access-token/{{access_token}}

This survey is a step towards critically assessing the 
way the scientific community measures the impact of 
its work.  We hope you will participate and help further 
this important research.

This survey is being conducted by Alexander Weissman and 
Filippo Radicchi from the School of Informatics at 
Indiana University. The research is sponsored by 
the NSF (Award #1446078) and approved by the Indiana 
University IRB (Protocol #1407480218). For more information, 
please write to Filippo Radicchi: filiradi@indiana.edu.

With regards,

Alexander Weissman and Filippo Radicchi
School of Informatics and Computing
Indiana University
                \end{verbatim}
            \end{center}
    \end{mdframed}
\caption{{\bf Email message for participation in the survey.}
	We sent this message to all potential participants in our survey.
	The email message contained a unique token key
	for every individual participant, pointing to
	her/his customized survey.
	}
\label{figLetter}
\end{figure}

\begin{figure}
\begin{center}
\includegraphics[width = 0.9\textwidth]{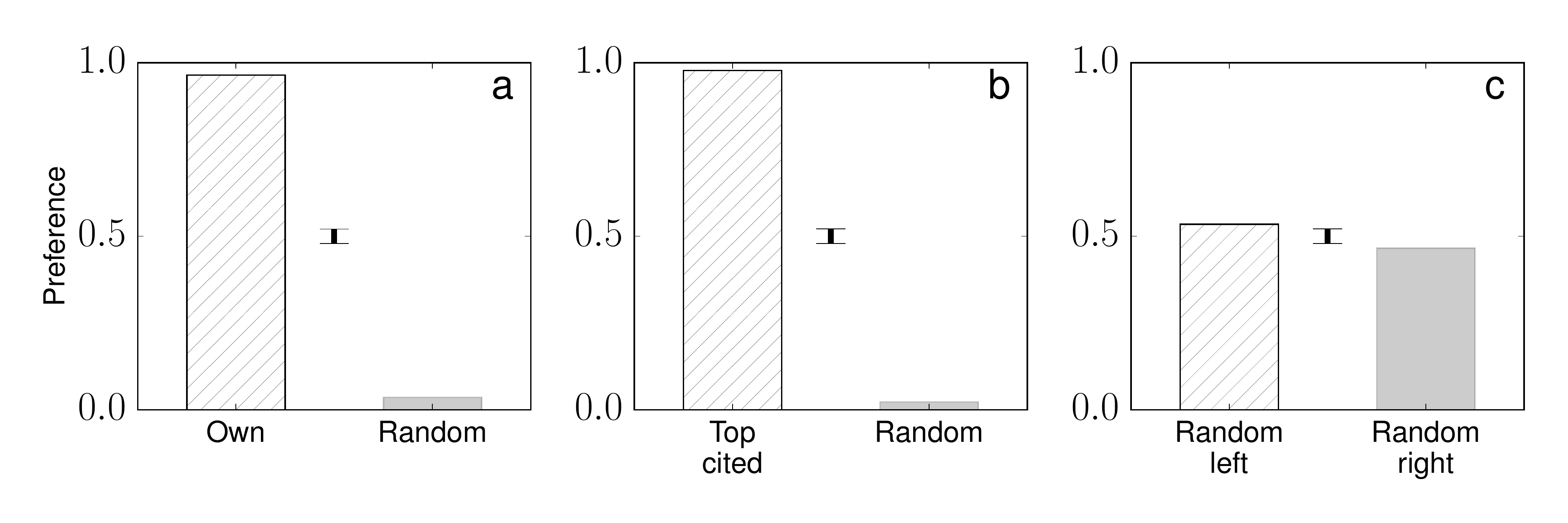}
\end{center}
\caption{
\noindent {\bf Sensitivity analysis.}
Results of this figure are based on the experiment
we conducted at Indiana University, the only
experiment where papers from the Random pool were considered.
{\bf a} Probability $P (a \to b | p_a=o, p_b=r)$  that the respondent preferred
the paper from the Own pool instead of the paper
from the Random pool. Such a probability
is $P (a \to b | p_a=o, p_b=r) = 0.97$ measured over $N = 587$ total comparisons, and leading to
a standard score $|z| = 22.50$.
{\bf b}
Same as in panel A but for comparisons between
one paper from the Top cited pool and the
other from the Random pool. In this case, we have:
$P (a \to b | p_a=t, p_b=r) = 0.98$, $N = 578$, and $|z| = 22.96$.
{\bf c} Results for comparisons
where both papers were taken from the Random pool.
We measure the probability $P (a \to b | g_a=L, g_b=R)$ that the respondent selected
the paper appearing on the left ($g_x = L$) or right ($g_x = R$)  
side of
the screen during the survey. We have: $P = 0.53$, 
$N = 550$, and $|z|=1.62$.
}
\label{figSM1}
\end{figure}

\clearpage

\begin{figure}
\begin{center}
\includegraphics[width = 0.9\textwidth]{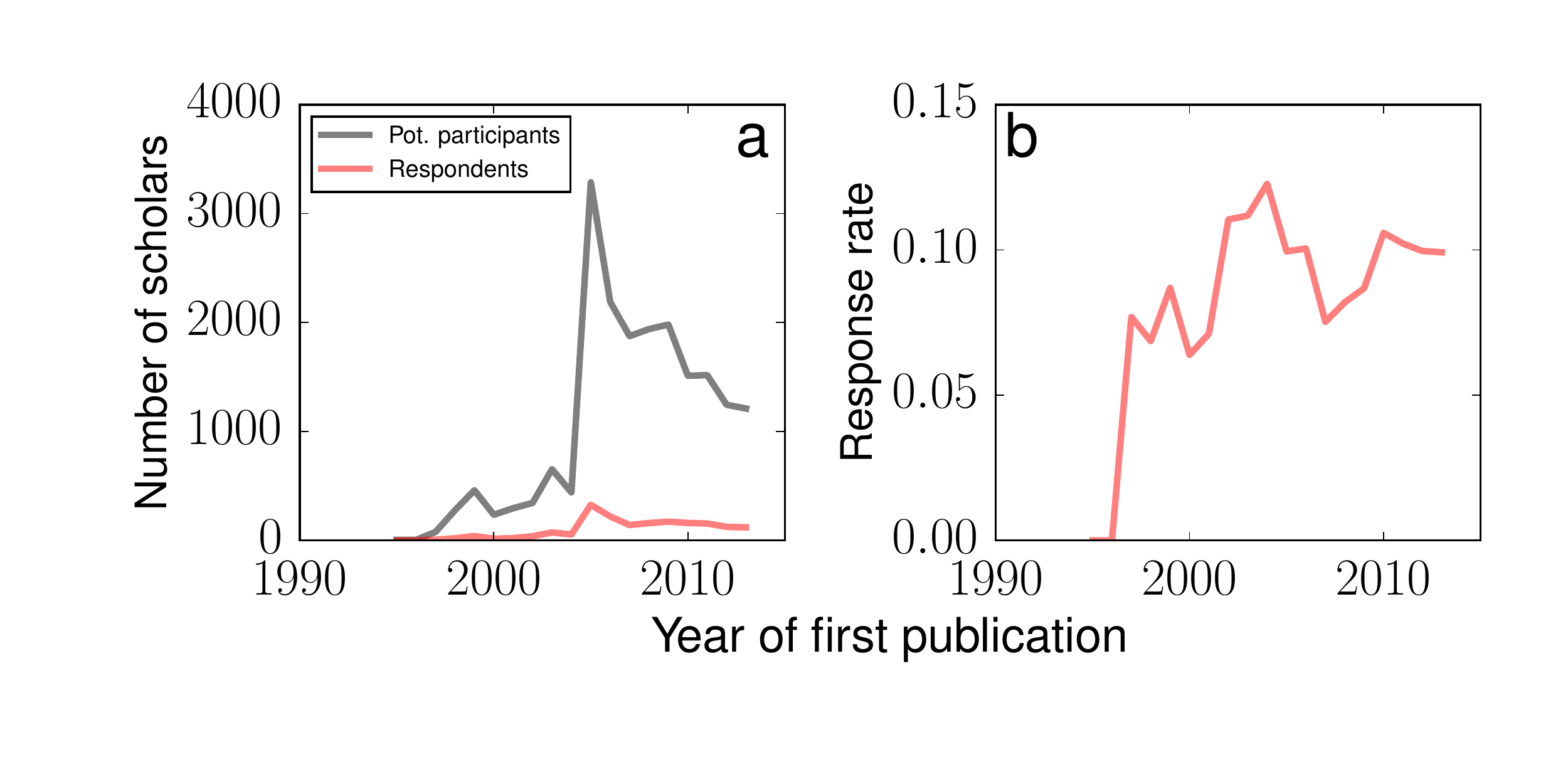}
\end{center}
\caption{
\noindent {\bf Participation in the survey.}
{\bf a} Total number of potential participants (gray line)
and respondents (red line) to the survey 
as functions of the year of their first publication
appearing in our database. The first publication corresponds
to the oldest paper where the email address 
of the scholar was appearing 
in the article metadata.
{\bf b}
Response rate (i.e., number of
respondents divided by number of potential participants)
as a function of the year of their first publication in the
database.
}
\label{figSM2}
\end{figure}

\begin{table}
    \centering
    \begin{tabular}{r r r r}
    Institution
         & 
         Potential participants
         &  
         Respondents
         &  
         Response rate \\
         \hline
         \hline
         Indiana University &
         $2,673$ &
         $313$ &
         $11.7\%$
        \\
        University of Michigan &
        $9,560$ &
        $889$ &
        $9.3\%$ 
        \\
        University of Minnesota &
        $7,313$ &
        $617$ &
        $8.4\%$
        \\
        \hline
        \hline
        &
        $19,546$ & 
        $1,840$ & 
        $9.4\%$ 
    \end{tabular}
    \caption{{\bf Participation in the survey.}
    Summary table listing number of potential participants, number of 
    respondents, and response rate for individual institutions.
    In the the last row of the table we list values of the same 
    quantities for the entire survey.
    }
    \label{tabSM1}
\end{table}

\begin{figure}
\begin{center}
\includegraphics[width = 0.9\textwidth]{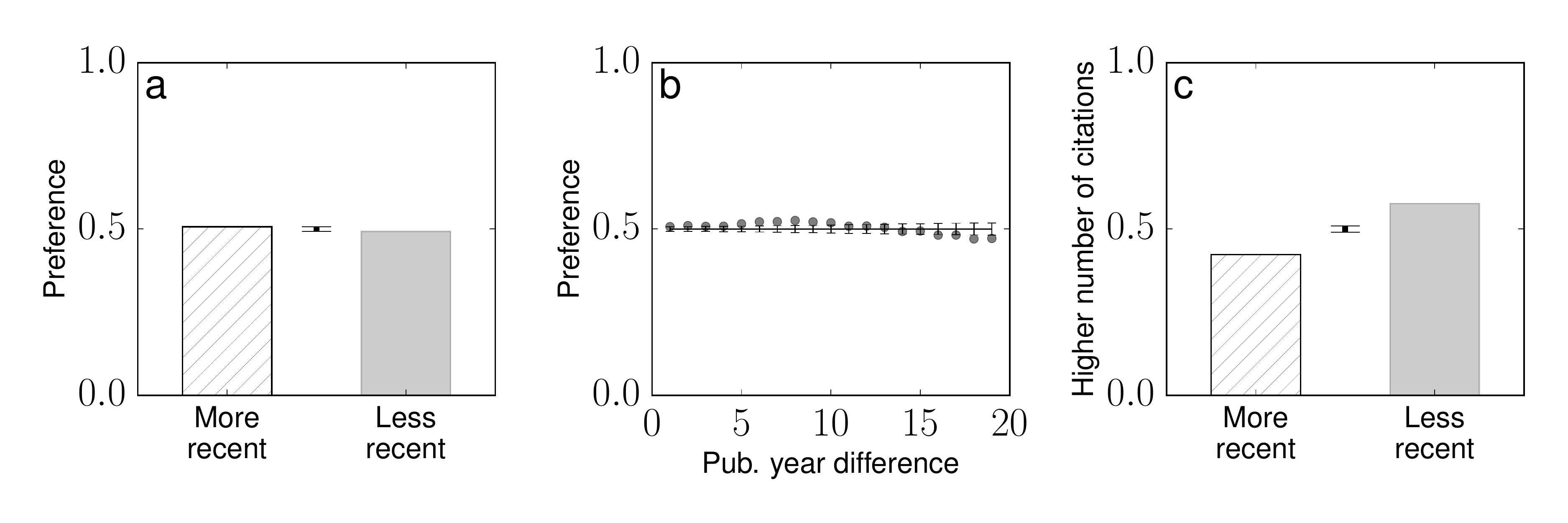}
\end{center}
\caption{
\noindent {\bf Perceived impact of top-cited publications depending on the age of papers.}
We consider only comparisons between
two papers taken from the ``Top-cited'' pool as in Fig.~\ref{fig2} of the main text.
{\bf a} Probability $P_{tt}(a \to b | y_a > y_b)$ that the respondent preferred
the more/less recent paper ($y_x$ is the year of publication
of paper $x$). To estimate the probability, we consider only
comparisons between papers with different 
years of publication, so that their difference
is different from zero.
More recent papers are preferred with
probability $P_{tt}(a \to b | y_a > y_b) = 0.51$
calculated over a sample 
of size $N = 5,609$, leading to a standard score
$|z| =  1.05$. Error bar is centered at $0.5$ and has height
equal to twice the standard deviation $\sigma$ of the 
unbiased binomial model. Here $\sigma = 0.007$. 
{\bf b}
Probability $P_{tt}(a \to b| y_a > y_b, y_a-y_b \geq \Delta y)$ that the respondent preferred
the more recent paper among the two in the
comparison as a function
of the difference of year of publication of the two papers. 
The horizontal line indicates the naive expectation of $0.5$
of the unbiased binomial model.
Error bars stand for standard deviation of the
binomial model.
{\bf c} Probability $P_{tt}(c_a > c_b| y_a > y_b)$ that the more/less recent
paper in the comparison has accumulated more citations.
To estimate the probability, we considered only
comparisons between papers with different 
years of publication and different citation numbers, so 
that the difference of both these numbers 
is different from zero. The probability is $P_{tt}(c_a > c_b| y_a > y_b) =0.42$, calculated over a sample
of size $N = 2,742$, leading to a standard score $|z| = 8.02$.
Standard deviation computed according to the
unbiased binomial model is $\sigma = 0.009$.
}
\label{figSM3A}
\end{figure}

\begin{figure}
\begin{center}
\includegraphics[width = 0.9\textwidth]{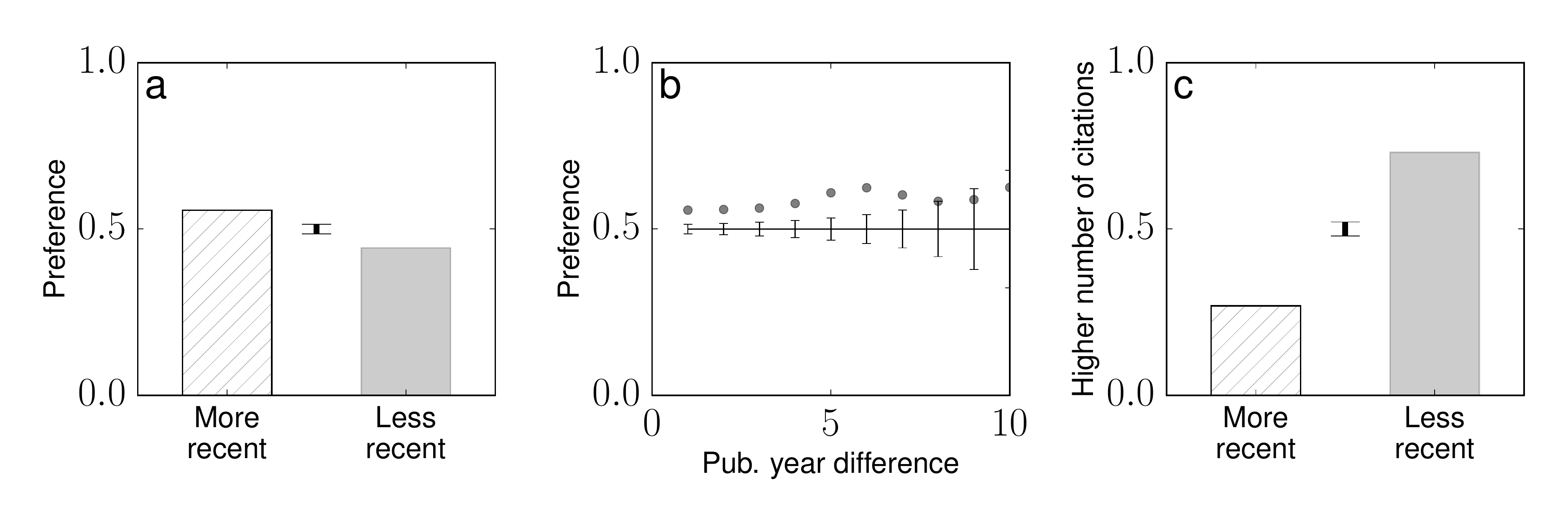}
\end{center}
\caption{
\noindent {\bf Perceived impact of authored publications depending on the age of papers.}
We consider only comparisons between
two papers taken from the ``Own'' pool as in Fig.~\ref{fig3} of the main text.
{\bf a} Probability $P_{oo}(a \to b | y_a > y_b)$ that the respondent preferred
the more/less recent paper ($y_x$ is the year of publication
of paper $x$). To estimate the probability, we consider only
comparisons between papers with different 
years of publication, so that their difference
is different from zero.
More recent papers are preferred with
probability $P_{oo}(a \to b | y_a > y_b) = 0.56$
calculated over a sample 
of size $N = 1,207$, leading to a standard score
$|z| =  3.94$. Error bar is centered at $0.5$ and has height
equal to twice the standard deviation $\sigma$ of the 
unbiased binomial model. Here $\sigma = 0.014$. 
{\bf b}
Probability $P_{oo}(a \to b| y_a > y_b, y_a-y_b \geq \Delta y)$ that the respondent preferred
the more recent paper among the two in the
comparison as a function
of the difference of year of publication of the two papers. 
The horizontal line indicates the naive expectation of $0.5$
of the unbiased binomial model.
Error bars stand for standard deviation of the
binomial model.
{\bf c} Probability $P_{oo}(c_a > c_b| y_a > y_b)$ that the more/less recent
paper in the comparison has accumulated more citations.
To estimate the probability, we considered only
comparisons between papers with different 
years of publication and different citation numbers, so 
that the difference of both these numbers 
is different from zero. The probability is $P_{oo}(c_a > c_b| y_a > y_b) =0.27$, calculated over a sample
of size $N = 550$, leading to a standard score $|z| = 10.83$.
Standard deviation computed according to the
unbiased binomial model is $\sigma = 0.021$.
}
\label{figSM3}
\end{figure}

\end{document}